# Observation of the Orbital Rashba-Edelstein Magnetoresistance

*Shilei Ding[1], Zhongyu Liang[1], Dongwook Go[2,3], Chao Yun[1], Mingzhu Xue[1], Zhou Liu[1], Sven Becker[3], Wenyun Yang[1], Honglin Du[1], Changsheng Wang[1], Yingchang Yang[1], Gerhard Jakob[3], Mathias Kläui[3,4,5], Yuriy Mokrousov[2,3],*

*Jinbo Yang[1,6,7]\**.

[1]State Key Laboratory for Mesoscopic Physics, School of Physics, Peking University, Beijing 100871, P.R. China.
[2]Peter Grünberg Institut and Institute for Advanced Simulation, Forschungszentrum Jülich and JARA, 52425 Jülich, Germany.
[3]Institute of Physics, Johannes Gutenberg-University Mainz, Staudingerweg 7, 55128 Mainz, Germany.
[4]Graduate School of Excellence Materials Science in Mainz, 55128 Mainz, Germany.
[5]Center for Quantum Spintronics, Department of Physics, Norwegian University of Science and Technology, NO-7491 Trondheim, Norway.
[6]Collaborative Innovation Center of Quantum Matter, Beijing, 100871, P.R. China.
[7]Beijing Key Laboratory for Magnetoelectric Materials and Devices, Beijing 100871, P. R. China.
*Correspondence to [jbyang@pku.edu.cn]

**We report the observation of magnetoresistance (MR) originating from the orbital angular momentum transport (OAM) in a Permalloy (Py) / oxidized Cu (Cu\*) heterostructure: the orbital Rashba-Edelstein magnetoresistance. The angular dependence of the MR depends on the relative angle between the induced OAM and the magnetization in a similar fashion as the spin Hall magnetoresistance (SMR). Despite the absence of elements with large spin-orbit coupling, we find a sizable MR ratio, which is in contrast to the conventional SMR which requires heavy elements. By varying the thickness of the Cu\* layer, we confirm that the interface is responsible for the MR, suggesting that the orbital Rashba-Edelstein effect is responsible for the generation of the OAM. Through Py thickness-dependence studies, we find that the *effective* values for the spin diffusion and spin dephasing lengths of Py are significantly larger than the values measured in Py / Pt bilayers, approximately by the factor of 2 and 4, respectively. This implies that another mechanism beyond the conventional spin-based scenario is responsible for the MR observed in Py / Cu\* structures – originated in a sizeable transport of OAM. Our findings not only unambiguously demonstrate current-induced torques without using any heavy elements via the OAM channel but also provide an important clue towards the microscopic understanding of the role that OAM transport can play for magnetization dynamics.**

The spin-orbit coupling (SOC) plays a critical role in the field of spin-orbitronics, leading to various mechanisms for electrical generation of spin such as bulk spin Hall effect (SHE) [1,2] and interfacial Rashba-Edelstein effect (REE) [3,4,5]. In the scenario of the SHE or REE, a pure spin current or non-equilibrium spin accumulation, whose spin polarization is transverse to the charge current, can be generated in materials with strong bulk SOC or at the interface with the Rashba-type SOC. Vice versa, a charge current can be generated from the spin current via the inverse SHE or inverse REE, which can be used for the electrical detection of spin [2,6-10]. In the past years, it has been uncovered that the interplay of the SHE (REE) and the inverse SHE (inverse REE) leads to an intriguing flavor of magnetoresistance (MR), commonly referred to as the spin Hall magnetoresistance (SMR) (Rashba-Edelstein magnetoresistance) [3,11-13]. The physics behind the SMR and Rashba-Edelstein magnetoresistance is the spin current reflection and absorption via spin-orbit torques (SOTs). It is often assumed that this spin-related MR requires strong SOC [11,12]. Within this paradigm, however, one would not expect that a light metal like Cu with negligible SOC could play a significant role in the generation of SOTs or SMR.

Recently, it has been shown that the orbital angular momentum (OAM) can be electrically induced at the surface or interface [14-19] via a process called the orbital Rashba-Edelstein effect (OREE) [15,20-22]. Despite its similarity to the spin REE, the OREE is independent of SOC and the inversion symmetry breaking alone is sufficient for its emergence. If the SOC is taken into consideration, the chiral orbital angular momentum texture couples to the spin angular momentum, leading to the coexistence of the spin REE and OREE [19]. Importantly, recent experiments indicate that the natural oxidation of Cu can lead to large SOTs [23-26], with the interpretation of the results agreeing that the physics of the OREE most likely plays a crucial role in the Cu oxide heterostructures, as the electrical generation of orbital current does not require SOC which is very weak in $CuO_x$ [25-27]. The surface of the naturally oxidized Cu is electrically insulating and the interface between the metallic ferromagnetic layer (FM) / $CuO_x$ is considered vital for the effect [24]. Indeed, the work of Ref. [26] indicates the emergence of the orbital current and orbital torque. The signs for the effective SOT are opposite in Py / Cu / $CuO_x$ and Fe / Cu / $CuO_x$, which cannot be explained by the conventional spin current scenario. One of the proposed mechanisms is that the OAM can be electrically induced by the OREE at the surface of the $CuO_x$, and the OAM is injected into the adjacent FM, where the OAM interacts with the local magnetic moment via the SOC [19,28]. In principle, a combination of the so-called orbital torque and its reciprocal process can results in a SMR-like effect, which originates in the OAM rather than the spin. However, to date, there is still a lack of report about the OREE-induced MR, which we refer to as the orbital Rashba-Edelstein MR (OREMR).

In this letter, we report the observation of the OREMR in the Py / oxidized Cu (denoted by Cu* hereafter) bilayer system. The angular dependence of the MR depends on the relative angle between the induced OAM and the magnetization similar to the SMR mechanism that presents for heavy metals with large SOC. In our systems with no elements with large SOC, we study the MR ratio, and by varying the thickness of the Cu* layer we check the bulk and interface contributions to the OAM. Through Py thickness dependence studies, we can ascertain the *effective* values for the spin diffusion and spin dephasing lengths of Py. From this we can deduce whether spin transport alone plays a role thus identifying to what extent OAM is responsible for the observed results.

All samples are prepared by pulsed laser deposition and the Hall bar devices with a line width of 75 μm are achieved during the deposition process with the help of a shadow mask. The fabricated Cu films are kept in the air for 48 h before transport measurements to obtain naturally oxidized Cu films (Cu*). The 3 nm Cu* is electrically insulating indicating the full oxidization of Cu. On the other hand, the 7 nm Cu* becomes conducting, which means that the oxygen concentration gradient is within 7 nm and there exists a Cu / $CuO_x$ mixture for the 7 nm Cu*. As a metallic FM layer, we use Py ($Ni_{81}Fe_{19}$). Details on the deposition conditions and shadow mask can be found in the Supplemental Material [29]. The magneto-transport measurements are performed at room temperature in a Quantum Design PPMS-9 cryostat which allows applying magnetic fields up to 9 T and rotating the samples in different planes with the specific sample holders.

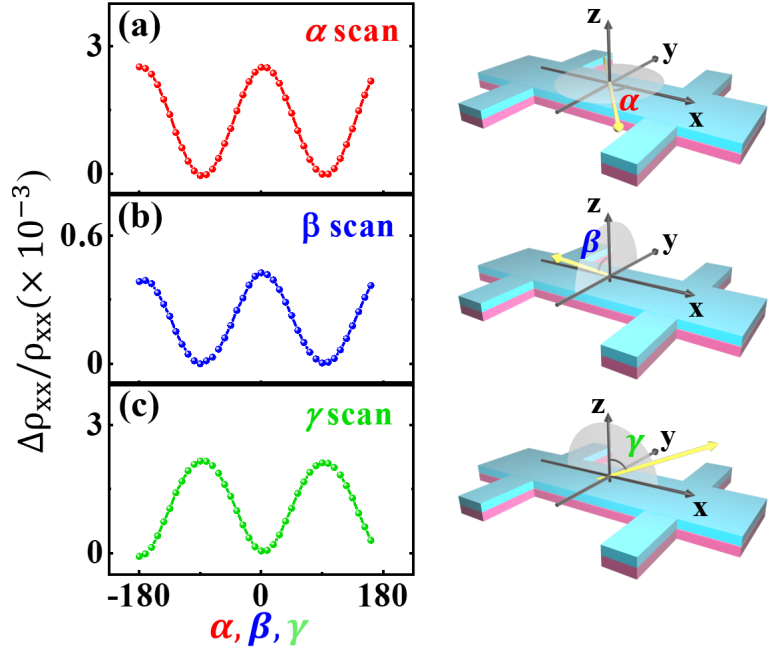

FIG. 1. (a)-(c) The angular dependent MR measurements in Py(5) / Cu*(3) heterostructure at 300 K and 6 T in the three $\mu_0 H$-rotation planes ($\alpha, \beta, \gamma$). The schematics on the right show the sample Hall bar and the definition of the axes, angles, and measurement configuration. $\Delta\rho_{xx}/\rho_{xx}$ is the MR ratio collected in different $\mu_0 H$-rotation planes.

We first measure the angular dependence of the MR in Py(5) / Cu*(3) sample by rotating the direction of the applied field $\mu_0 H$ in $xy$ plane ($\alpha$ scan), $yz$ plane ($\beta$ scan), and $zx$ plane ($\gamma$ scan), where $\alpha$, $\beta$, and $\gamma$, are the angles between $\mu_0 H$ and $x$, $z$, and $z$ axes, respectively. The applied field strength is set to 6 T, which is large enough to saturate the magnetization ($M_s$) along the $\mu_0 H$ direction. As shown in Fig. 1, we observe a sizable MR in all three orthogonal planes. Note that the MRs for the $\alpha$ and $\gamma$ scans are nearly an order of magnitude larger than the MR for the $\beta$ scan, indicating different microscopic mechanisms. The $\alpha$ and $\gamma$ scan MR ratios are consistent with the prediction of the anisotropic MR (AMR) of the Py layer, where the Cu*(3) = $CuO_x$(3) is insulating, and the applied charge current flows in the Py layer. Note that $\alpha = 0°$ and $\gamma = 90°$ correspond to the identical physical situation with magnetic field and current directions parallel. The AMR phenomenology of polycrystalline FM predicts $\rho_{xx} = \rho_\perp + (\rho_\| - \rho_\perp) m_x^2$ [34], where $\rho_\|$ and $\rho_\perp$ are the resistivity when the magnetization direction $\boldsymbol{m}$ aligns along and perpendicular to the charge current direction, $m_x$ is the $x$ component of $\boldsymbol{m} = \boldsymbol{M}/M_s$. This predicts that the variations of the resistivity in the $\alpha$ and $\gamma$ scans follow $\frac{\Delta \rho_{xx}(\alpha)}{\rho_{xx}(\alpha)} \sim cos^2 \alpha$ and $\frac{\Delta \rho_{xx}(\gamma)}{\rho_{xx}(\gamma)} \sim sin^2 \gamma$, which are consistent with the data shown in Figs. 1(a) and 1(c), respectively.

However, the AMR cannot explain the MR obtained in the $\beta$ scan, where we find $\frac{\Delta \rho_{xx}(\beta)}{\rho_{xx}(\beta)} \sim cos^2 \beta$ as shown in Fig. 1(b). The observed MR in the $\beta$ scan exhibits similar angular dependence as the SMR scenario, where the resistivity is given by $\rho_{xx} = \rho_0 - \rho_1 m_y^2$. Here $\rho_0$ is the resistivity offset, $\rho_1$ is the magnitude of the resistivity change, and $m_y$ is the $y$ component of $\boldsymbol{m}$. Considering the negligible SOC of Cu*, a conventional SHE-induced SMR cannot lead to the novel MR in the $\beta$ scan. Also, the spin REE is believed not to be the origin of SOT in Py / Cu* heterostructures [26], and thus the Rashba-Edelstein MR cannot explain the result. From the above considerations, we argue that the OREMR is the most plausible mechanism for the observed MR in the $\beta$ scan. At the interface of Py (5) / Cu* (3), the inversion symmetry is broken and the orbital asymmetry leads to the Rashba-type texture of the OAM [19]. The applied electric field can induce a finite OAM through the OREE. In this case, the induced OAM at Py / Cu* interface points along the $\boldsymbol{E} \times \hat{\boldsymbol{z}}$ direction, which is analogous to the interfacial mechanism for the generation of the spin [35] but does not require SOC. As the interface-generated OAM is injected into the FM, it interacts with the local magnetic moment via the SOC and $s$-$d$ exchange interaction, which is called the orbital torque mechanism [28]. We remark that the angular dependences of the orbital torque and the spin-injection torque are qualitatively similar, and so are the angular dependences of the OREMR and SMR. We also note that the MR for the $\alpha$ scan also contains the contribution from the OREMR but the signal is dominated by the AMR by an order of magnitude.

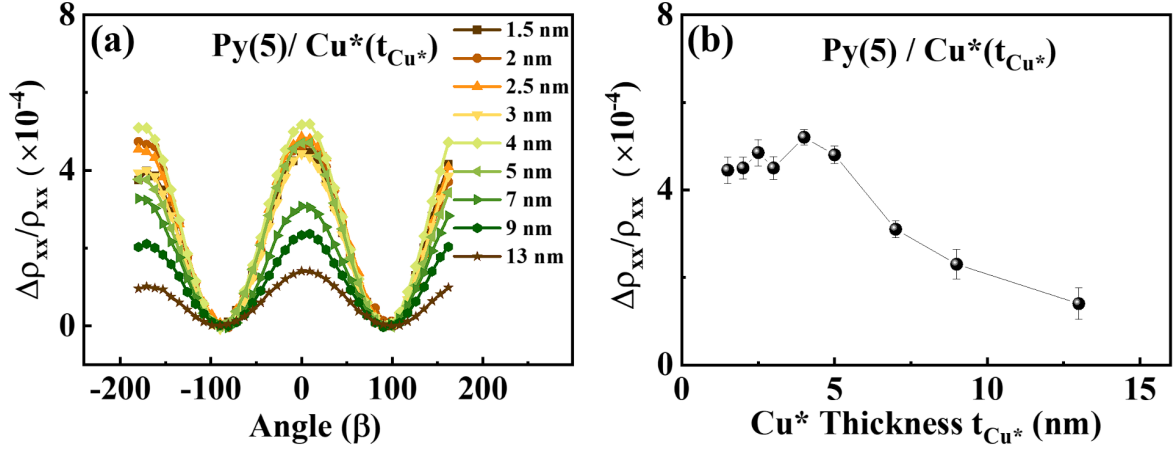

FIG. 2. (a) Angular dependence of the OREMR for Py (5) / Cu* ($t_{Cu^*}$) samples with different Cu* layer thicknesses. (b) OREMR ratio $\frac{\Delta\rho_{xx}}{\rho_{xx}}$ as a function of the thickness of Cu*. The MR ratio keeps nearly constant for $t_{Cu^*} \leq 5$ nm, indicating a typical interfacial mechanism. For $t_{Cu^*} > 5$ nm, the MR ratio decreases monotonically as the thickness of Cu* increases.

To further understand the properties of the observed OREMR, we carry out Cu* thickness-dependent measurements. Figure 2 (a) presents the $\beta$ scan MR of Py (5) / Cu* ($t_{Cu^*}$) for various Cu* thicknesses $t_{Cu^*}$. Note that while the magnitudes differ, all the samples follow the SMR-like angular dependence $\frac{\Delta\rho_{xx}(\beta)}{\rho_{xx}(\beta)} \sim cos^2\beta$. We extract the MR ratio by $\frac{\Delta\rho_{xx}}{\rho_{xx}} = [\rho_{xx}(\beta = 0º) - \rho_{xx}(\beta = 90º)]/\rho_{xx}(\beta = 90º)$ for each sample, showing the $t_{Cu^*}$ dependence of the MR ratio in Fig. 2(b). When the thickness of $t_{Cu^*}$ is below 5 nm, the naturally oxidized Cu is electrically insulating and the MR ratio stays nearly constant, highlighting that the OREMR originates in the Py / Cu* interface. We find that even for Py (5) / Cu* (1.5) we do not observe any decrease of the MR compared to other samples. For thicker Cu* samples ($t_{Cu^*} > 5$ nm), we find a monotonic decrease of the MR as $t_{Cu^*}$ increases. While the monotonic decrease may be attributed to the current shunting through the non-oxidized region of Cu*, we remark that it could also be due to intrinsic quenching of the OAM since the pure Cu without oxidization has negligible $d$ orbital character near the Fermi energy [19]. We note that a similar behavior was also observed in the measurement of the orbital torque in AlO$_x$/Cu/CoFe, where the decrease of the torque efficiency seems to be much more drastic than the current shunting effect [27].

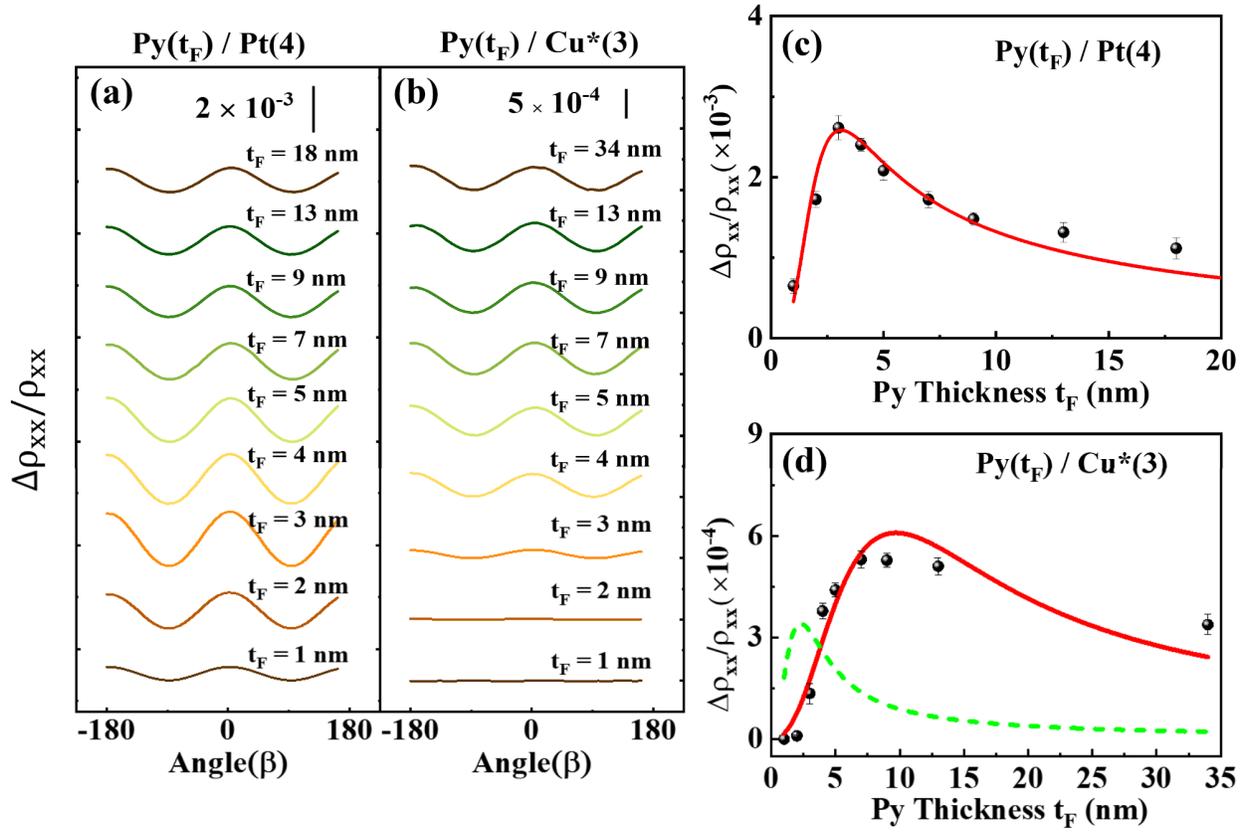

FIG. 3. (a) and (b) show the OREMR measurements in the $\beta$ scan for Py($t_F$) / Pt(4) and Py($t_F$) / Cu*(3) performed at 6 T. (c) and (d) demonstrate the Py thickness dependence of the MR ratio $\frac{\Delta\rho_{xx}}{\rho_{xx}}$ taken from (a) and (b), respectively. (c) The SMR ratio for Pt (4) / Py ($t_F$) reaches the peak at 3 nm and then decreases rapidly with increasing Py thickness due to the current shunting. (d) On the other hand, the OREMR ratio for Cu* (3) / Py ($t_F$) is negligible for $t_F < 2$ nm and increases slowly, reaching the maximum value at 7 nm. Also, the decrease of the OREMR for $t_F > 15$ nm by the current shunting is occurring on a longer length scale. The red lines are the fitting curves of the data. The Green dashed line in (d) is the fitting by forcing the parameters of Py to values taken from the Py / Pt samples, which shows a significant deviation from the data.

To identify the OREMR as the origin and compare it to the conventional SMR, we next study the FM thickness dependence of the OREMR in comparison with the SMR in Py / Pt. The angular dependence of the SMR in Py($t_F$) / Pt(4) and the OREMR Py($t_F$) / Cu*(3) are shown in Figs. 3(a) and 3(b), respectively, for various thicknesses of $t_F$. From these data, we plot the thickness dependences of the SMR in Py($t_F$) / Pt(4) and the OREMR Py($t_F$) / Cu*(3) as a function of $t_F$ in Fig. 3(c) and 3(d), respectively. For Py($t_F$) / Pt(4) samples, the SMR ratio exhibits a peak at ~ 3 nm and decreases for thicker Py samples. While the increase of the SMR for

$t_F < 3$ nm is due to the spin dephasing in Py, the decrease for thicker Py is due to current shunting. We note that our result is similar to the previous studies [36,37]. For the quantitative analysis, we use the diffusion model developed by Kim et al. [13], which takes into account the charge current shunting and the *longitudinal* spin current diffusing into the FM in magnetic bilayers. According to this model, the SMR is given by

$$\frac{\Delta\rho_{xx}}{\rho_{xx}} \sim \theta_{\text{eff}}^2 \frac{\lambda_N}{t_N} \frac{\tanh^2(t_N/2\lambda_N)}{1+\xi} \times \left[\frac{g_R}{1+g_R\coth(t_N/2\lambda_N)} - \frac{g_F}{1+g_F\coth(t_N/2\lambda_N)}\right],$$

$$g_R = 2\rho_N\lambda_N\text{Re}[G_{\text{mix}}],$$

$$g_F = (1-P^2)\frac{\rho_N\lambda_N}{\rho_F\lambda_F}\tanh(t_F/\lambda_F), \quad (1)$$

where $t_N$, $\rho_N$, $\lambda_N$, and $\theta_{\text{eff}}$ represent the thickness, resistivity, spin diffusion length, and the effective spin Hall angle of nonmagnetic metal, $G_{\text{mix}}$ is the spin-mixing conductance at the interface, $\rho_F$, $\lambda_F$, and $P$ refer to the resistivity, spin diffusion length, and current spin polarization of the FM layer, respectively. Here, $\xi = \rho_N t_F/\rho_F t_N$ reflects the current shunting effect. We note that the total torque exerted by the spin current on the Py layer increases and saturates with the increase of the thickness of Py. However, for a Py layer thinner than the spin dephasing length, the spin Hall angle exhibits a smaller value because the *transverse* spin current is not sufficiently absorbed by FM, which leads to a smaller SMR ratio [38]. The spin dephasing effect is described by [39]

$$\theta_{\text{eff}} = \theta_{\text{SH}}[1-\text{sech}(k_F t_F)] \quad (2)$$

where $\theta_{\text{SH}}$ is the saturated spin Hall angle, $k_F = \sqrt{\lambda_F^{-2} + \lambda_\phi^{-2}}$, where $\lambda_\phi$ is the spin dephasing length. The fitting of the thickness dependence of the SMR is shown with a red curve in Fig. 3(c). For simplicity, we assume a spin transparent interface, where the conductivity is limited by the Sharvin conductance, and assume $\text{Re}[G_{\text{mix}}] = 0.68 \times 10^{15}\ \Omega^{-1}\text{m}^{-2}$ for the Py / Pt interface [40]. Other parameters used in the fitting are the following: $\rho_N = 1.2 \times 10^{-6}\ \Omega\text{m}$, $\lambda_N = 1.6$ nm, $\lambda_F = 4.5$ nm, $\lambda_\phi = 1.3$ nm, and $P = 0.49$ for Py [41]. The thickness dependence of $\rho_F$ can be found in the Supplemental Material [29]. We obtain an effective spin Hall angle for Pt, $\theta_{\text{SH}} = 0.208$, which corresponds to the spin Hall conductivity $\sigma_{\text{SH}} = 1.7 \times 10^5\ (\hbar/2e)(\Omega \cdot m)^{-1}$.

The overall behavior of the Py thickness dependence of the OREMR in Py ($t_{FM}$) / Cu* (3) is similar to that of the SMR in Py($t_F$) / Pt(4). However, the maximum value appears at $t_F = 7$ nm, which is much larger than the value measured in Py($t_F$) / Pt(4). Also, the decrease by the current shunting effect is less drastic. For the analysis, we develop a diffusion model for interface/FM systems, whose derivation is found in the Supplemental Material [30]. It is given by

$$\frac{\Delta\rho_{xx}}{\rho_{xx}} \sim \lambda_{\text{int}}^2[1-\text{sech}(k_F t_F)]^2 \left[\frac{2\rho_F\lambda_F\text{Re}[G_{\text{mix}}]-(1-P^2)\tanh(t_F/\lambda_F)}{\lambda_F t_F}\right], \quad (3)$$

where $\lambda_{\text{int}}$ describes the efficiency of the spin current generation from the interface, which has the dimension of a length. We note that the model assumes only the spin degree of freedom without taking the orbital degree of freedom into account, so the parameters in Eq. (3) should be understood as effective values. For the fitting in Fig. 3(d), we assume $\text{Re}[G_{\text{mix}}] = 0.68 \times 10^{15}\ \Omega^{-1}\text{m}^{-2}$ [40] and $P = 0.49$ [41], the rest of the parameters are optimized from the fitting, where we find $\lambda_F = 8.64$ nm, $\lambda_\phi = 5.3$ nm, and $\lambda_{\text{int}} = 0.087$ nm. Remarkably, $\lambda_F = 8.64$ nm, $\lambda_\phi = 5.3$ nm obtained in Py ($t_F$) / Cu* (3) are much larger than $\lambda_F = 4.5$ nm, $\lambda_\phi = 1.3$ nm, the values obtained in Py($t_F$) / Pt(4). If the spin current was responsible for the observed MR in Py/ Cu*, the parameters for Py should be comparable for the two different studied systems. If we force the parameters to be $\lambda_F = 4.5$ nm and $\lambda_\phi = 1.3$ nm for the fitting, the fitting curve deviates qualitatively from the measured data. This confirms that the conclusion that the origin of the MR in Py ($t_F$) / Cu* (3) is not the convensional spin mechanism due to the SHE or spin REE, and indicates that the MR is rooted in the OAM dynamics.

Finally, we discuss the difference between the conventional SMR and the OREMR mechanisms. The SMR in Py / Pt bilayers arises as a result of the combined effect of the spin Hall torque and its reciprocal effect. The reflection of the spin current depends on the relative angle between the magnetization and the spin polarization of the spin current, which leads to the modulation of the longitudinal resistivity for different angles of the magnetization. For thin FMs, the SMR increases as the FM thickness increases until the FM thickness becomes comparable to the spin dephasing length. Further increasing the thickness of the FM reduces the magnitude of the SMR due to charge current shunting. On the other hand, for the OREMR in Py / Cu*, the OAM generated by the OREE at the interface is injected into the FM and interacts with the local magnetic moment *indirectly* through the SOC. The modulation of the longitudinal resistivity depends on the relative angle between the direction of the magnetization and the OAM of the conduction electron. This leads to the same angular dependence as that of the SMR at the end. The main difference between the SMR and OREMR lies in the microscopic processes in the FM. Since the ways in which spin and OAM interact with the local magnetic moments are fundamentally different, it is natural to expect different *effective* spin diffusion and dephasing lengths between the spin-based mechanism and the OAM-based mechanism. We speculate that the small energy scale of the SOC compared to the *sd* exchange interaction may result in large values of the effective diffusion and dephasing lengths. This implies that the coupling of the OAM and the local magnetic moment or the orbital-to-spin conversion occurs over a larger length scale than the spin dephasing length. However, verification of the microscopic mechanism requires further theoretical investigation, which goes beyond the scope of this work.

In conclusion, we report a magnetoresistance effect in Py / naturally-oxidized Cu bilayers at room temperature. This effect cannot be explained by conventional SMR due to the absence of any heavy element with large SOC, and we term this OREMR. By varying the thickness of the oxidized Cu layer, we find that the mechanism originates from the interface, indicating OREE. Careful examination of the Py thickness dependence shows that the spin diffusion and spin dephasing lengths significantly deviate from the values measured in conventional Py / Pt bilayers. This strongly suggests that the conventional spin mechanism cannot serve as the microscopic origin of the MR observed in Py / naturally-oxidized Cu bilayers, supporting the OREMR scenario. Our findings not only unambiguously demonstrate the current-induced torque without using any heavy element via the OAM channel but also provide an important clue towards the microscopic understanding of the interaction of the out-of-equilibrium OAM and magnetization in magnetic materials. The results of our work point to an exciting possibility that low-spin orbit materials can be efficiently used for the transport of orbital angular momentum with much higher efficiency than that associated with the transport of spin, which might prove to be crucial in taking the next step from conventional spintronics to orbitronics that can be realized without expensive and environmentally harmful heavy metal materials.

## Acknowledgments

We acknowledge the support from the National Key Research and Development Program of China (Grant No. 2017YFA0206303, 2016YFB0700901, 2017YFA0403701). National Natural Science Foundation of China (Grant No. 51731001, 11675006, 11805006, 11975035), Graduate School of Excellence Materials Science in Mainz (MAINZ), Deutsche Forschungsgemeinschaft (DFG, German Research Foundation) Spin+X (A01, A11, B02) TRR 173 – 268565370 and Project No. 358671374. The work was further supported by the Horizon 2020 Framework Programme of the European Commission under FETOpen Grant Agreement No. 863155 (s-Nebula) and the European Research Council Grant Agreement No. 856538 (3D MAGiC).

S. Ding, Z. Liang, and D.Go contribute equally to this work.

field dependence of OREMR ratio for Py (3) / Cu* (3) and Py (4) / Cu* (3), the field dependence of OREMR for Py($t_F$) / Pt(4) and Py($t_F$) / Cu*(3), MR in $yz$ plane scan for the single Py layer, the field dependence of OREMR for Py(5) / Cu*(3), the fitting details for the Py thickness dependence of OREMR ratio, and the resistivity for Py($t_F$) / Pt(4) and Py($t_F$) / Cu*(3), which includes Refs.[13, 30-33, 39-41].